\documentclass[aps,preprint,a4]{revtex4}

\usepackage{amsmath}
\usepackage{bm}

\newcommand{\be}{\begin{equation}}
\newcommand{\ee}{\end{equation}}
\newcommand{\bea}{\begin{eqnarray}}
\newcommand{\eea}{\end{eqnarray}}
\newcommand{\bsube}{\begin{subequations}}
\newcommand{\esube}{\end{subequations}}

\newcommand{\stab}[1]{\begin{tabular}[c]{c}#1\end{tabular}}
\newcommand{\doeps}[3]{\stab{\setlength{\epsfxsize}{#1}%
   \setlength{\epsfysize}{#2}\epsffile{#3}}}
\usepackage{epsfig}

\begin{document}

\title{Theory of proton coupled electron transfer reactions: 
Assessing the Born-Oppenheimer approximation for the proton motion
using an analytically solvable model}
\author{Renhui Zheng}
\author{Yuanyuan Jing}
\author{Liping Chen}
\author{Qiang Shi}\email{qshi@iccas.ac.cn}
\affiliation{Beijing National Laboratory for
Molecular Sciences, State Key Laboratory for Structural Chemistry of
Unstable and Stable Species, Institute of Chemistry, Chinese Academy
of Sciences, Zhongguancun, Beijing 100190, China}

\begin{abstract}

By employing an analytically solvable model including
the Duschinsky rotation effect,
we investigated the applicability of 
the commonly used Born-Oppenheimer (BO) approximation
for separating the proton and proton donor-acceptor 
motions in theories of proton coupled electron 
transfer (PCET) reactions.
Comparison with theories based on the BO approximation shows that,
the BO approximation for the proton 
coordinate is generally valid while some further approximations
may become inaccurate in certain range of parameters.
We have also investigated the effect of vibrationally coherent tunneling
in the case of small reorganization energy, and shown that it plays 
an important role on the rate constant and kinetic isotope effect. 
\end{abstract}
\maketitle
\section{introduction}
\label{sec:intro}
Proton coupled electron transfer (PCET) reactions are important 
in many chemical, biological and electro-chemical 
processes.\cite{mayer04,tommos98,costentin06,huynh07,reece09} 
We consider in this paper the case of 
concerted PCET reactions, where the 
proton and electron are transferred simultaneously with no
stable intermediates. 
Theories of PCET reactions were developed over the past two 
decades. Cukier has proposed a theory for concerted PCET based on 
nonadiabatic transitions between multiple vibronic 
states,\cite{cukier94,cukier96,cukier98} which treats the proton 
coordinate as a high frequency intramolecular vibrational mode,
and leads to rate constant expressions similar to
the Bixon-Jortner model\cite{jortner76}  for the electron 
transfer (ET) reactions.\cite{marcus64,marcus85}
Hammes-schiffer and coworkers have generalized this theory
to include the environmental collective coordinate that couples to 
the proton coordinate.\cite{soudackov99,soudackov00,hammes-schiffer01}

Recent works on the PCET theory have focused on the importance 
of the proton donor-acceptor motion. The main concept is that 
fluctuations of the donor-acceptor separation $R_{DA}$ strongly
affect the vibrational wave function overlap for the 
proton coordinate, and thus the effective transfer integral 
between vibronic states. Such effect has been first investigated
in the case of vibrationally nonadiabatic proton 
transfer reactions, and is found to play an 
important role and leads to much smaller kinetic isotope effect (KIE)
than simple estimation using the overlap of vibrational wave functions
at the equilibrium proton donor-acceptor 
distance.\cite{borgis91,suarez91a,kiefer04,kiefer04a}
The effect of the donor-acceptor motion has also been 
widely discussed in recent PCET theories.\cite{knapp02,soudackov05,
hatcher05,hammes-schiffer08,klinman06,cukier04,presse06}
In these theoretical treatments, the proton degree of freedom 
is first quantized, and a Born-Oppenheimer approximation is 
applied to separate the proton motion and the slower degrees of 
freedom that couple to it. 
The PCET problem is then reduced to transitions between 
a group of vibronic states. 
Within the above theoretical framework,
the effect of donor-acceptor motion can be treated either
statically by doing a thermal average over the $R_{DA}$ 
distribution,\cite{knapp02,klinman06,cukier04} or can be treated  
dynamically using a time correlation function 
formalism.\cite{soudackov05,hatcher05,hammes-schiffer08,presse06}

It is interesting to investigate the applicability of 
the BO approximation to the proton motion for several reasons.
First, although the masses of proton and deuterium are much less than 
those of the heavy atoms, the difference is not as dramatic 
as the mass ratio between the electron and nuclei. When the 
hydrogen bond is stiff between the donor and acceptor, high frequency 
donor-acceptor motion raise the question whether a separation of 
time scales is still appropriate. 
Second, even after the BO approximation is applied to separate the
proton motion and the proton donor-acceptor motion,
additional approximations such as static donor-acceptor 
motion,\cite{knapp02,klinman06,cukier04}
or analytical approximations on the vibrational wave 
function overlap are often employed.\cite{soudackov05,hammes-schiffer08,
presse06} 
It is also desirable to test these approximations 
quantitatively using a solvable model.

In this paper, we provide tests of the above mentioned
BO approximation for the proton motion using a model Hamiltonian 
for concerted PCET reactions. We show that, under certain well defined 
approximations, the effect of proton donor-acceptor motion
can be incorporated in a model Hamiltonian base on the 
Duschinsky rotation effect (DRE).\cite{duschinsky37} 
The DRE describes the mixing of the normal modes
between the donor and acceptor potential energy surfaces,\cite{duschinsky37} 
which is different from the linear 
displaced harmonic modes used in the conventional 
spin-boson model.\cite{leggett87,weiss99a}
During the past years, the effect of DRE have been widely 
discussed in areas such as
electronic spectroscopy,\cite{kubo55,yan86,egorov98a,
mebel99}  nonadiabatic relaxations,\cite{kubo55,
egorov99b,peng07} and
electron transfer reactions.\cite{lee00,sando01,tang03,liang03,velizhanin09} 
However, to the best of our knowledge, a model based on 
the DRE has not been applied in studies of PCET previously.

When the electronic coupling is small, the DRE model allows us to 
calculate the PCET reaction rates analytically using the Fermi's 
golden rule. This analytically solvable model is then applied 
to assess the BO approximation for the proton motion, as well as 
the static donor-acceptor motion approximation and the analytical 
approximations to the vibrational wave function 
overlaps. The remainder of this paper is arranged as follows.
The model Hamiltonian and theories to calculate PCET rate constants
are presented in Sec. \ref{sec:theory}. The results are presented 
in Sec. \ref{sec:results}, where PCET rates and the KIEs 
are calculated using the FGR, and compared with results obtained 
from approximate theories. 
We have also studied the rate constant and KIE in the case of 
small electronic reorganization energy 
happen,\cite{presse06,hatcher07,edwards09,ludlow09}
with the focus on the possible role of vibrationally coherent 
tunneling.\cite{onuchic88,topaler96}
The main conclusions are summarized in Sec. \ref{sec:conclusions}.

\section{Theory}
\label{sec:theory}
\subsection{Model Hamiltonian}
We consider the case of concerted proton and electron
transfer with a small electronic coupling (nonadiabatic ET), where 
a two electronic surfaces description 
is sufficient,\cite{hammes-schiffer01,hammes-schiffer08} 
and the Fermi's golden rule (FGR) that treats the electronic coupling
to first order perturbation is valid.\cite{weiss99a}
To this end, we apply an extended spin-boson model to describe the
PCET reactions. The total Hamiltonian can be written as
\begin{equation}
{H} = {H}_S + {H}_B + {H}_{BS} \; ,
\label{eq:modelh}
\end{equation}
where, the system Hamiltonian $H_S$ includes terms
related to the electronic degree of freedom (DOF),
the proton reaction coordinate $x$, and 
the proton donor-acceptor separation $R_{DA}$,
\begin{equation}
{H}_S = \hbar \Delta {\sigma}_x 
+ \frac{\Delta G}{2} {\sigma}_z 
+ H_a | a \rangle \langle a | + H_b | b \rangle \langle b | 
~~,
\label{HS}
\end{equation}
where $|a\rangle$ and $|b\rangle$ denote the electron donor
and acceptor states, ${\sigma}_x = 
|a \rangle \langle b| + |b \rangle \langle a|$,
and ${\sigma}_z = |b \rangle \langle b| - |a \rangle \langle a|$.
$H_a$ and $H_b$ are defined as
\begin{equation}
\label{eq:haandb}
H_{a,b} = \frac{p^2}{2m_H}  + \frac{P^2}{2M} 
+ \frac{1}{2}m_H\omega_H^2
\left(x\pm \frac{d}{2}\pm \kappa\frac{R}{2} \right)^2 
+ \frac{1}{2}M\Omega^2 R^2 \;.
\end{equation}
Here, $p$ and $x$, $P$ and $R$ are the momenta and coordinates
of the proton and proton donor-acceptor DOFs, respectively;
$m_H$ and $\omega_H$, $M$ and $\Omega$ are the corresponding mass 
and frequencies;
$R$ is defined as the difference between the donor-acceptor separation
and its equilibrium value $R\equiv R_{DA} - R_{DA}^{eq}$;
$\kappa$ is the coupling coefficient between the proton equilibrium 
position and the proton donor-acceptor separation $R$, and
$\pm(d+\kappa R)/2$ is the equilibrium position of the proton coordinate
on the donor and acceptor surfaces. 

The potential energy surface presented in Eq. (\ref{eq:haandb}) 
leads to shifted proton equilibrium positions on the donor and
acceptor surfaces as a function of the proton 
donor-acceptor distance $R$. As the $x$ and 
$R$ motions are coupled differently on the $|a\rangle$ and $|b\rangle$
surfaces, there is a rotation of the corresponding normal modes 
(see also Fig. \ref{fig1}),
which is the DRE introduced in the previous Sec. \ref{sec:intro}.
In the case of a widely used 
linear model for the proton and the electron/proton
donors and acceptors,\cite{knapp02,soudackov05,hatcher05, hatcher05a,presse06}
$\kappa = 1$.
Without loss of generality, we will assume such case and 
drop $\kappa$ in the following derivations.

We now briefly discuss the relevance of the DRE in PCET reactions.
The V-shaped $x$-$R$ potential energy surfaces for the donor and acceptor
states as shown in Fig. 1 can actually be found in many previous
publications studying proton transfer (PT) (e.g., Fig. 1 in Ref.\cite{pu06}) 
and PCET reactions (e.g., Fig. 4 in Ref.\cite{hatcher04}), as an 
effect that the 
proton equilibrium positions shift in different direction
on the donor and acceptor surfaces as a function of the proton 
donor-acceptor distance. 
Under the harmonic approximation, these potential energy surfaces
display a rotation of the normal modes that can be
described by the DRE.
The parameters for a DRE model can thus be obtained by
analyzing the donor and acceptor 
potential energy surfaces.

We also assume that the solvent DOFs
couple only to the electronic DOF, and the
bath Hamiltonian can be written as
\begin{equation}
H_B = \sum_{j=1}^N\left(\frac{p_j^2}{2m_j} + \frac{1}{2}m_j \omega_j^2 x_j^2
\right) \; ,
\end{equation}
where ${x}_j$, ${p}_j $, $m_j$, $\omega_j$ are the coordinates, momenta,
masses, and frequencies of the harmonic bath modes. 

The bath DOFs are assumed to couple linearly with the 
electronic DOF, and {the system-bath coupling} term is given by
\begin{equation}
{H}_{BS} = - \sum_{j=1}^N c_j x_j \sigma_z \;\;.
\label{eq:vbs}
\end{equation}
The system-bath coupling is usually characterized using the 
spectral density $J(\omega)$ defined as
\begin{equation}
\label{eq:sdensity} J(\omega) = \frac{\pi}{2}\sum_{j=1}^{N}
\frac{c_{j}^2}{
 m_j\omega_{j}} \delta(\omega - \omega_{j}) \;\; .
\end{equation}

In general, the proton DOF also couples to the 
environmental DOFs, which leads to 
a reorganization energy associated with proton 
motion,\cite{soudackov99,soudackov00,hammes-schiffer01}
as well as vibrational energy relaxation and 
dephasing.\cite{oxtoby81,owrutsky94}
Although we did not consider such coupling in this paper,
it can certainly be incorporated by extending the
above model Hamiltonian described 
in Eqs. (\ref{eq:modelh}-\ref{eq:vbs}).

\subsection{Fermi's golden rule}
In the electronic nonadibatic limit where $\hbar\Delta$ is small,
first order perturbation can be applied, and
the rate constant can be calculated using Fermi's 
Golden Rule:\cite{weiss99a}
\begin{equation}
\label{eq:k10}
k = \Delta^2 \int_{-\infty}^{\infty}
e^{-i\frac{\Delta G}{\hbar}t} C(t) dt \; ,
\end{equation}
where the correlation function $C(t)$ is defined as
\begin{equation}
C(t) = \frac{1}{Z_0} {\rm Tr} \left[ e^{-\beta H_0} 
e^{\frac{i}{\hbar} H_0 t}
e^{-\frac{i}{\hbar} H_1 t} \right] \;\; .
\end{equation}
Here, $Z_0$ is the partition function of the donor state,
$Z_0 = {\rm Tr} e^{-\beta H_0}$, 
\begin{equation}
H_0 = H_a + H_B + \sum_j c_j x_j \; ,
\end{equation}
and 
\begin{equation}
H_1 = H_b + H_B - \sum_j c_j x_j \; .
\end{equation}

Since the proton and proton donor-acceptor DOFs are decoupled from 
the bath modes, $C(t)$ can be calculated as
\begin{equation}
\label{eq:ctotal1}
C(t) = \frac{1}{Z_a} {\rm Tr} \left[ e^{-\beta H_a} 
e^{\frac{i}{\hbar} H_a t}
e^{-\frac{i}{\hbar} H_b t} \right]C_B(t) \;\; .
\end{equation}

$C_B(t)$ on the right hand side of the above Eq. (\ref{eq:ctotal1})
arises from the bath contribution and is defined as
\begin{equation}
\label{eq:cbt}
C_B(t) = \frac{1}{{\rm Tr} e^{-\beta H_{B0} }} 
{\rm Tr} \left[ e^{-\beta H_{B0}} 
e^{\frac{i}{\hbar} H_{B0} t}
e^{-\frac{i}{\hbar} H_{B1}  t} \right]\; ,
\end{equation}
where $H_{B0,B1} = H_B \pm \sum_j c_j x_j$.
It can be calculated analytically using the spectral density $J(\omega)$,
resulting in the following equation:\cite{weiss99a}
\begin{eqnarray}
C_B(t) 
& =  & \exp \left \{ -\frac{4}{\hbar}\int_{0}^\infty 
d\omega \frac{J(\omega)}{\pi\omega^2}\left[
\coth(\beta\hbar\omega/2)(1-\cos \omega t) 
+ i \sin \omega t \right]
\right\} \; .
\end{eqnarray}

As in the many previous theories for the PCET reactions,
we apply the following approximation for $C_B(t)$, 
which can usually be obtained by a high temperature approximation
and short-time expansion:
\begin{equation}
\label{eq:cbath}
C_B(t) \approx  e^{-\frac{\lambda t^2}{\beta\hbar^2} 
- i \frac{\lambda}{\hbar}t} \; ,
\end{equation}
where $\lambda$ is the electronic reorganization energy
\begin{equation}
\lambda = \int_{0}^\infty d\omega \frac{4J(\omega)}{\pi\omega} \; .
\end{equation} 

The first term on the right hand side of the above Eq. (\ref{eq:ctotal1}),
$\frac{1}{Z_a} {\rm Tr} \left[ e^{-\beta H_a} 
e^{\frac{i}{\hbar} H_a t}
e^{-\frac{i}{\hbar} H_b t} \right]$
describes the time correlation function of a two-mode DRE model, and can
be calculated as  
\begin{equation}
\label{eq:ctxR}
\frac{1}{Z_a} {\rm Tr} \left[ e^{-\beta H_a} 
e^{\frac{i}{\hbar} H_a t}
e^{-\frac{i}{\hbar} H_b t} \right] = \frac{1}{Z_a}\int dxdRdx'dR'
\langle xR | e^{\frac{i}{\hbar} H_a (t+i\hbar\beta)} | x'R'\rangle
\langle x'R' | e^{-\frac{i}{\hbar} H_b t} | xR\rangle \; .
\end{equation}
Since $H_a$ and $H_b$ are Hamiltonians of harmonic oscillators,
the matrix elements can be calculated analytically,\cite{feynman65}
and the correlation function can then be obtained
using Gaussian integrals. Details
of such calculation can be found in many previous publications
such as Refs.\cite{kubo55,egorov98a,peng07,velizhanin09}, and will
not be presented in this paper. 

We further note that the harmonic $x$-$R$ model was employed in 
this study in order to obtain an analytical expression for the 
time correlation function in Eq. (\ref{eq:ctotal1}). Including the 
anharmonic effects numerically in some of the widely used linear models (e.g.,
Refs. \cite{hatcher05,hatcher07,edwards09}) would also be 
straightforward, as calculating 
Eq. (\ref{eq:ctxR}) in such cases only need to solve a two-dimensional
Schr\"{o}dinger equation, which can be done routinely on modern
computers. 
We note that a similar idea of solving a two-dimensional
Schr\"{o}dinger equation for the proton and proton accepter-donor motion 
was proposed in Ref.\cite{soudackov05}.

\subsection{Rate constant within the BO approximation for the proton motion}
Due to the small mass of proton or deuterium, a
BO approximation is often applied to separate
the motion of the proton coordinates $x$ and the proton 
donor-acceptor separation
$R$. For a fixed donor-acceptor separation $R$, the proton coordinate
can be quantized, and the vibronic 
states can be calculated as
\begin{equation}
\label{eq:vibronic}
\left[ \frac{p^2}{2m_H} + V_{\alpha}(x,R) \right]
|\phi_{j,\alpha} (x,R) \rangle
= E_{j,\alpha}|\phi_{j,\alpha } (x,R) \rangle \; ,
\end{equation}  
where $\alpha = a, b$ denotes the electronic donor and acceptor states,
and $j$ denotes the vibrational states for the proton DOF. 
According to Eq. (\ref{eq:haandb}), 
$V_{a,b}(x,R) = 1/2m_H \omega_H^2 (x\pm d/2 \pm R/2)^2$, and
$E_{j,\alpha}$ and $|\phi_{j,\alpha } (x,R) \rangle$ can be obtained from
the eigenstates of displaced harmonic-oscillators.

In the case of small electronic coupling considered 
in the above subsection, neglecting the non-BO coupling terms
between the vibronic states in Eq. (\ref{eq:vibronic}) will 
formulate the PCET reaction
as nonadiabtic transitions between a group of vibronic 
states.\cite{cukier94,cukier98,hammes-schiffer01,
hammes-schiffer08}
By further assuming an initial equilibrium distribution on 
the donor vibrational states, 
the total rate constant can be calculated as
\begin{equation}
\label{eq:ktot}
k = \sum_{\mu\nu} P_{\mu,a}  k_{\mu\nu} \; ,
\end{equation}
where $P_{\mu,a}$ is the equilibrium population of vibrational 
state $|\mu\rangle$ on the donor potential energy surface;
and $k_{\mu\nu}$ is the transition rate constant from state 
$|\mu,a\rangle$ to $|\nu,b\rangle$.
The rate constant $k_{\mu\nu}$ are then calculated
using the effective system Hamiltonian involving states 
$|\mu, a \rangle$  and $|\nu, b\rangle$\cite{cukier94,cukier98,
hammes-schiffer01, hammes-schiffer08}
\begin{eqnarray}
{H}^{\mu\nu}_S  & =  & \hbar \Delta \langle \phi_{\mu} | \phi_{\nu}\rangle 
\left( | \mu,a \rangle \langle \nu, b | 
 + c.c.\right)
+ \left[-{\Delta G}/{2} + (\mu+1/2)\hbar\omega_H\right] | \mu,a\rangle
\langle \mu,a |
\nonumber \\ & &
\left[ {\Delta G}/{2} + (\nu+1/2)\hbar\omega_H\right] | \nu, b\rangle
\langle \nu,b |
+ \frac{P^2}{2M} + \frac{1}{2}M\Omega^2 R^2 
~~.
\label{HS2}
\end{eqnarray}

Since fluctuations of the the donor-acceptor coordinate $R$
can cause changes of the overlap integral
$\langle \phi_{\mu}|\phi_{\nu}\rangle$ by orders of magnitudes, 
\cite{borgis91,suarez91a,
kiefer04,kiefer04a,soudackov05,hammes-schiffer08,knapp02,
klinman06,presse06}
there is no general exact analytical expression for $k_{\mu\nu}$, 
and additional 
approximations are often employed\cite{cukier94,cukier98,kuznetsov99,
hammes-schiffer01,kiefer04a,hammes-schiffer08}. We will briefly
present the results in applying two widely used approximations
to the model system presented in the previous subsections
II. A and B in the case of nonadiabatic ET reactions.

In the rate constant expression originally proposed by Kutnetsov 
and Ulstrup,\cite{kuznetsov99}
the $R$-mode is treated statically, and the rate constant 
is obtained as a thermally average over the classical 
Boltzmann distribution of $R$: 
\begin{equation}
\label{eq:kuk0}
k_{\mu\nu} \approx \Delta^2 
\sqrt\frac{\pi}{\lambda k_BT}
\exp\left[-\frac{(\Delta G_{\mu\nu}+\lambda)^2}
{4\lambda k_BT}\right] \int  d R P(R) |S_{\mu\nu}(R)|^2\; ,
\end{equation}
where $\Delta G_{\mu\nu} =  \Delta G + (\nu-\mu)\hbar\omega_H$,
$S_{\mu\nu}(R) = \langle \phi_{\mu} (R) | \phi_{\nu}(R)\rangle$ 
is the Franck-Condon overlap of the vibrational wave functions,
and $P(R) = \sqrt{M\Omega^2/2\pi k_B T} e^{-M\Omega^2R^2/2 k_B T}$.
This static approximation has recently been applied 
to PCET reactions in enzymes by Klinman and coworkers.\cite{knapp02,
klinman06}
For cases where $\hbar\Omega > k_B T$, the above Eq. (\ref{eq:kuk0})
can be extended to include the quantum effect of the $R$-mode by 
using a quantum mechanic distribution for the 
$R$-mode,\cite{kiefer04a,hammes-schiffer08a}
\begin{equation}
\label{eq:kuk1}
k_{\mu\nu} \approx \Delta^2 
\sqrt\frac{\pi}{\lambda k_BT}
\exp\left[-\frac{(\Delta G_{\mu\nu}+\lambda)^2}
{4\lambda k_BT}\right] \int  d R P_{qm}(R) |S_{\mu\nu}(R)|^2\; ,
\end{equation}
where 
$P_{qm}(R) = \sqrt{1/2\pi\langle R^2 \rangle} e^{-R^2/2\langle R^2 \rangle}$, 
and $\langle R^2 \rangle = {\hbar}/{2M\Omega}
\coth(\beta\hbar\Omega/2) $. This extended UK expression will be used
in later calculations for the static $R$-mode approximation.

Another approximation is to expand
$S_{\mu\nu}(R)$ around the equilibrium position $R=0$ using
an exponential function, which 
is widely used in PT\cite{borgis89,suarez91,kiefer04} 
and PCET\cite{soudackov05,hammes-schiffer08} theories:
\begin{equation}
\label{eq:expr}
S_{\mu\nu}(R) \approx S_{\mu\nu}(0)e^{-\alpha_{\mu\nu}R} \; .
\end{equation}
The problem is now equivalent to an extended 
spin-boson model with exponential coupling.\cite{nitzan75}
$k_{\mu\nu}$ can then be calculated as
\cite{soudackov05,hammes-schiffer08} 
\begin{equation}
\label{eq:kmunu}
k_{\mu\nu}= |\Delta S_{\mu\nu}(0)|^2
\int_{-\infty}^{\infty} dt \exp[\alpha_{\mu\nu}^2 (C_R(0) + C_R(t))]
 e^{-i\frac{\Delta G_{\mu\nu}}{\hbar} t -\frac{\lambda t^2}{\beta\hbar^2} 
- i \frac{\lambda}{\hbar}t}
\end{equation}
where 
\begin{equation}
\label{eq:crt}
C_R(t) =  \frac{\hbar}{2M\Omega}\left[
\coth(\beta\hbar\Omega/2)\cos \Omega t
- i \sin \Omega t \right]
\; .
\end{equation}

In later studies, we will denote Eqs. (\ref{eq:ktot}), (\ref{eq:kmunu}), 
and (\ref{eq:crt}), as the exponential coupling approximation.

\section{results}
\label{sec:results}

\subsection{Rate constant and KIE}
\begin{figure}
\begin{center}
\doeps{5in}{3.95in}{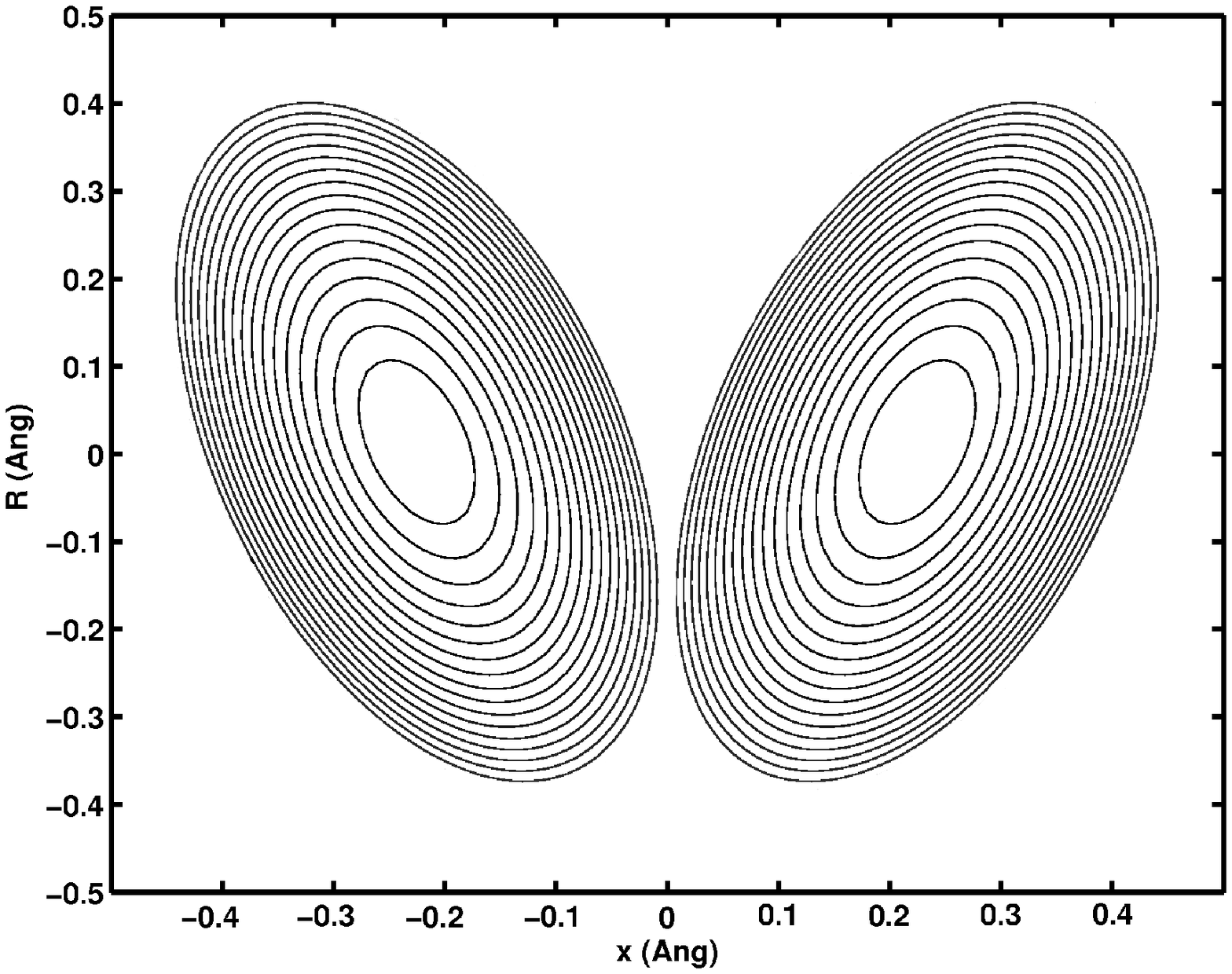}
\caption{Potential energy surfaces for the electronic donor (left) and 
acceptor (right) states.
%\vspace{5em}
}
\label{fig1}
\end{center}
\end{figure}

We now apply the above theories to calculate PCET rate 
constants and KIEs. 
Fig. \ref{fig1} shows the donor and acceptor potential energy surfaces 
from the Hamiltonian $H_{a}$ and $H_b$ defined in Eq. (\ref{eq:haandb}).
We will assume $\omega_H$ = 3000 cm$^{-1}$ throughout this paper,
the other parameters used in Fig. \ref{fig1} are $M$ = 100 amu, 
$\Omega = 100$ cm$^{-1}$, and $d = 0.45$ \AA. 
It can be seen that when $R$ is smaller than its equilibrium 
value $R=0$, the distance between the energy minima 
for the proton coordinate on the donor and acceptor surfaces 
becomes smaller, which will lead to enhanced proton tunnelling.
A key feature of the DRE, which is the rotation between normal 
modes on the donor and acceptor potential energy surfaces, 
can be clearly seen in Fig. \ref{fig1}.

\begin{figure}
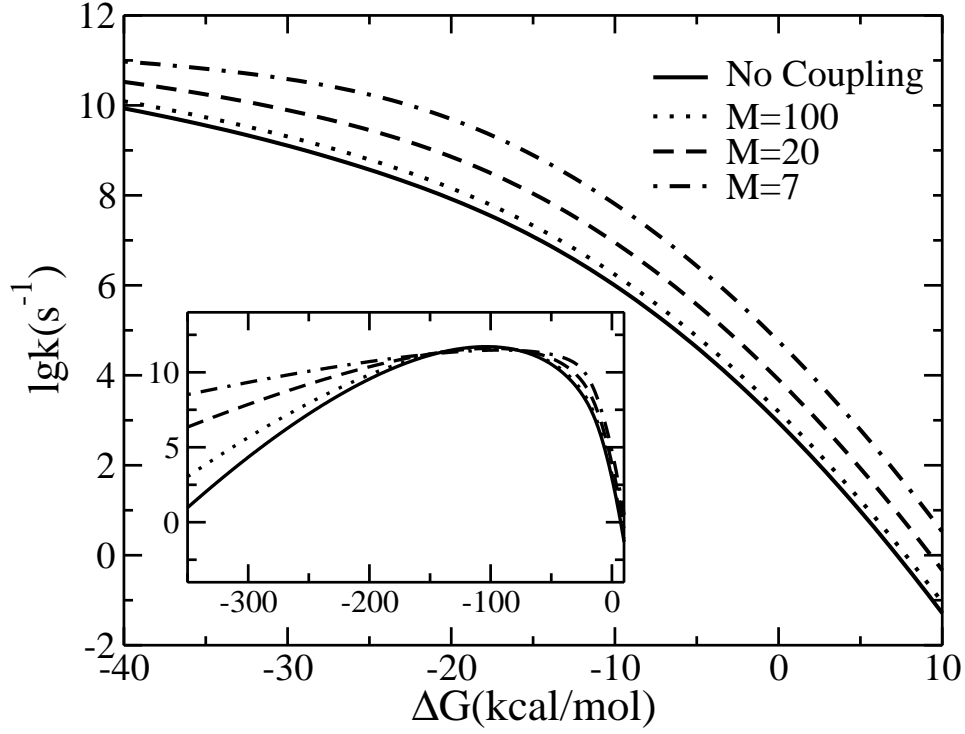

\begin{center}
\doeps{5in}{3.82in}{fig2}
\vspace{2em}
\caption{PCET rate constants as a function of $\Delta G$.
Curves from bottom to top: No coupling to the $R$-mode;
$M$ = 100 amu, $M$ = 20 amu, and
$M$ = 7 amu.  The other 
parameters are: $\lambda$ = 30 kcal/mol, $\Delta$ = 100 cm$^{-1}$,
$\Omega$ = 300 cm$^{-1}$,
$d$ = 0.45 \AA, and $T$ = 300 K.
}
%\vspace{5em}
\label{fig2}
\end{center}
\end{figure}

Fig. \ref{fig2} shows the PCET rate constant as a function of the 
driving force $\Delta G$ using the DRE model, where 
the proton coordinate $x$ is 
coupled to $R$-modes with different parameters.
$\lambda =$ 30 kcal/mol, $\Delta = 100$ cm$^{-1}$, 
$d$ = 0.45 \AA, and $T=300$ K were used in the calculation. 
It can be seen that coupling to the $R$-mode can significantly 
increase the rate constants by orders of magnitude,
especially when the donor-acceptor mass is small.
We also note that the PCET rate constants keep increasing
with more negative $\Delta G < -\lambda$. The reason is that 
the overall reorganization energy is very large
after taking account into the contribution from the proton 
coordinate, and a real turnover of the PCET rate can only happen 
at very negative $\Delta G$ (see the inset of Fig. \ref{fig2}). 
This effect was also observed in recent studies
by Hammes-Schiffer and coworkers,\cite{edwards09,edwards09a}
where the explanation is based on vibronic transitions to high
vibrational states.
Such high activation barrier for ET indicates that the 
proton transfer should happen mainly through quantum tunnelling.

The KIE is an important character of 
reactions involving proton transfer, which is defined as
the ratio of the rate constant for hydrogen to that for deuterium.
The dependence of PCET rate constants and KIEs on
various parameters has been widely discussed in previous theoretical
studies.\cite{cukier98,hammes-schiffer01,hammes-schiffer08,klinman06,cukier04}
More specifically, the effects of different donor-acceptor parameters
and driving forces have been investigated by Hammes-Schiffer and coworkers
recently.\cite{edwards09, edwards09a} 
The main purpose of the calculations below is, however,
to quantitatively assess the 
applicability of the BO approximation for the proton motion: the 
static $R$-mode and the exponential coupling approximations
will be evaluated
using the analytically solvable model Hamiltonian presented in Sec. II. 

\begin{figure}
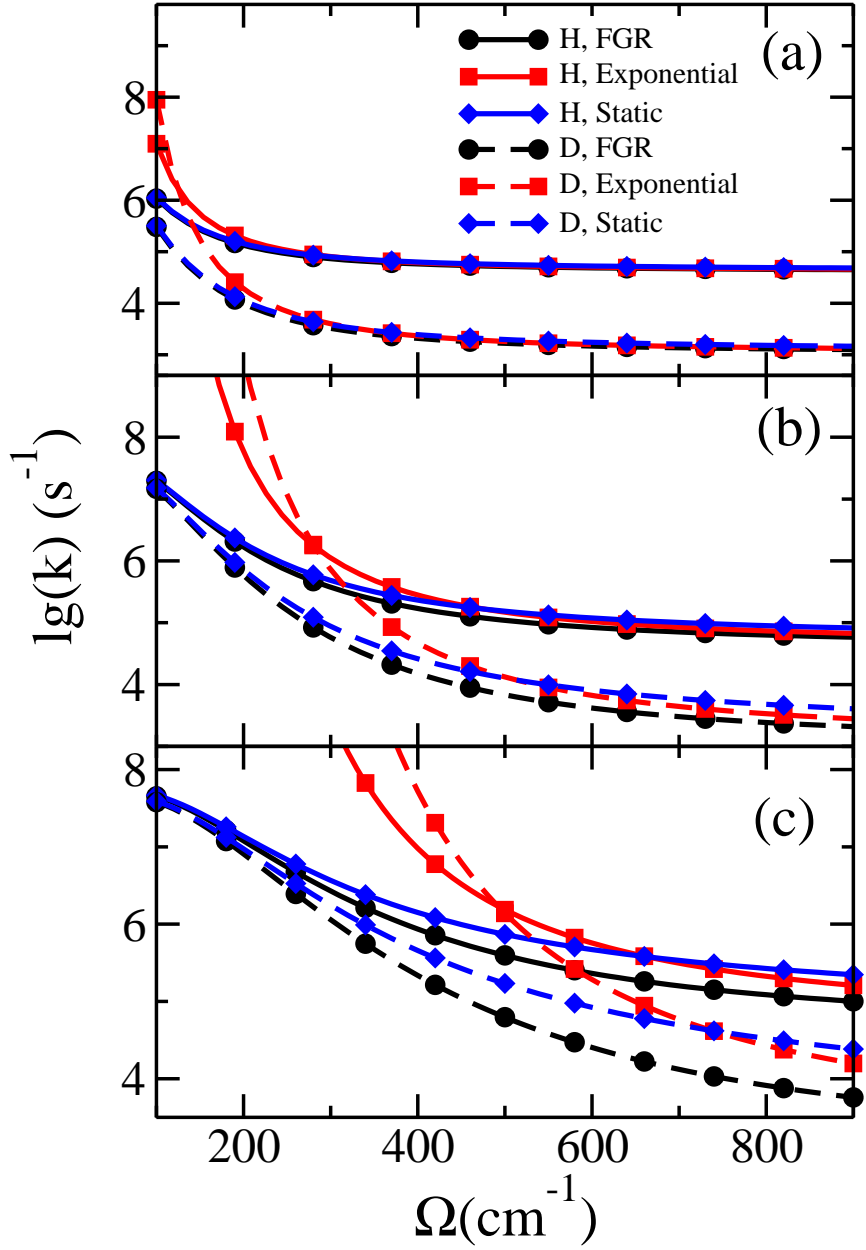

\begin{center}
\doeps{4.5in}{6.54in}{fig3}
%\vspace{2em}
\caption{Rate constants as a function of the proton donor-acceptor frequency
$\Omega$ for proton (H) and deuterium (D). (a) M = 100 amu, (b) M = 20 amu,
(c) M = 7 amu.
The other parameters are $\lambda$ = 30 kcal/mol, 
$\Delta G$ = -5 kcal/mol,
$\Delta$ = 100 cm$^{-1}$,
$d$ = 0.45 \AA, and $T$ = 300 K.
}
\label{fig3}
\end{center}
\end{figure}

Fig. \ref{fig3} plots the PCET reaction rate constants for proton
and deuterium as a function of the donor-acceptor motion 
frequency. Comparisons with the rate constant expression 
using exponential coupling approximation
(Eqs. \ref{eq:ktot}, \ref{eq:kmunu}, and
\ref{eq:crt}), and the extended UK expression within the 
static $R$-mode approximation (Eqs. \ref{eq:ktot}
and \ref{eq:kuk1}) are also shown. 
Three different donor-acceptor masses $M$ = 100 amu, 20 amu, and 7 amu were
considered. The other parameters are 
$\lambda = 30$ kcal/mol, $\Delta G$ = -5 kcal/mol, $d=$ 0.45 \AA, and 
$T=300$ K. It can be seen that the rate constant increases 
significantly with the decrease of the $R$-mode mass, where
larger proton donor-acceptor fluctuation leads to larger
enhancement of the PCET rates.  The extended UK
rate expression Eq. (\ref{eq:kuk1}) agrees well with the 
exact FGR result except for the small mass and high 
frequency cases ($M$=20 amu and $\Omega>$500 cm$^{-1}$, 
$M$=7 amu and $\Omega>$400 cm$^{-1}$), 
which indicates that the dynamical effect of the $R$-mode becomes
important only at high frequencies.
We can also see that the exponential coupling approximation
Eq. (\ref{eq:kmunu}) starts to overestimate the rate constants 
and becomes inaccurate for 
small mass and small donor-acceptor frequency
($\Omega < 200$ cm$^{-1}$ for $M$=100 amu, $\Omega < 400$ cm$^{-1}$ 
for $M$=20 amu, $\Omega < 700$ cm$^{-1}$ 
for $M$=7 amu), with a steep rise of the rate constants 
for low donor-acceptor frequency $\Omega$. 
The reason is that, the range of fluctuation 
for $R$ is quite large in such cases, and the exponential
approximation for the vibrational wave function overlap in 
Eq. (\ref{eq:expr}) becomes invalid. 

\begin{figure}
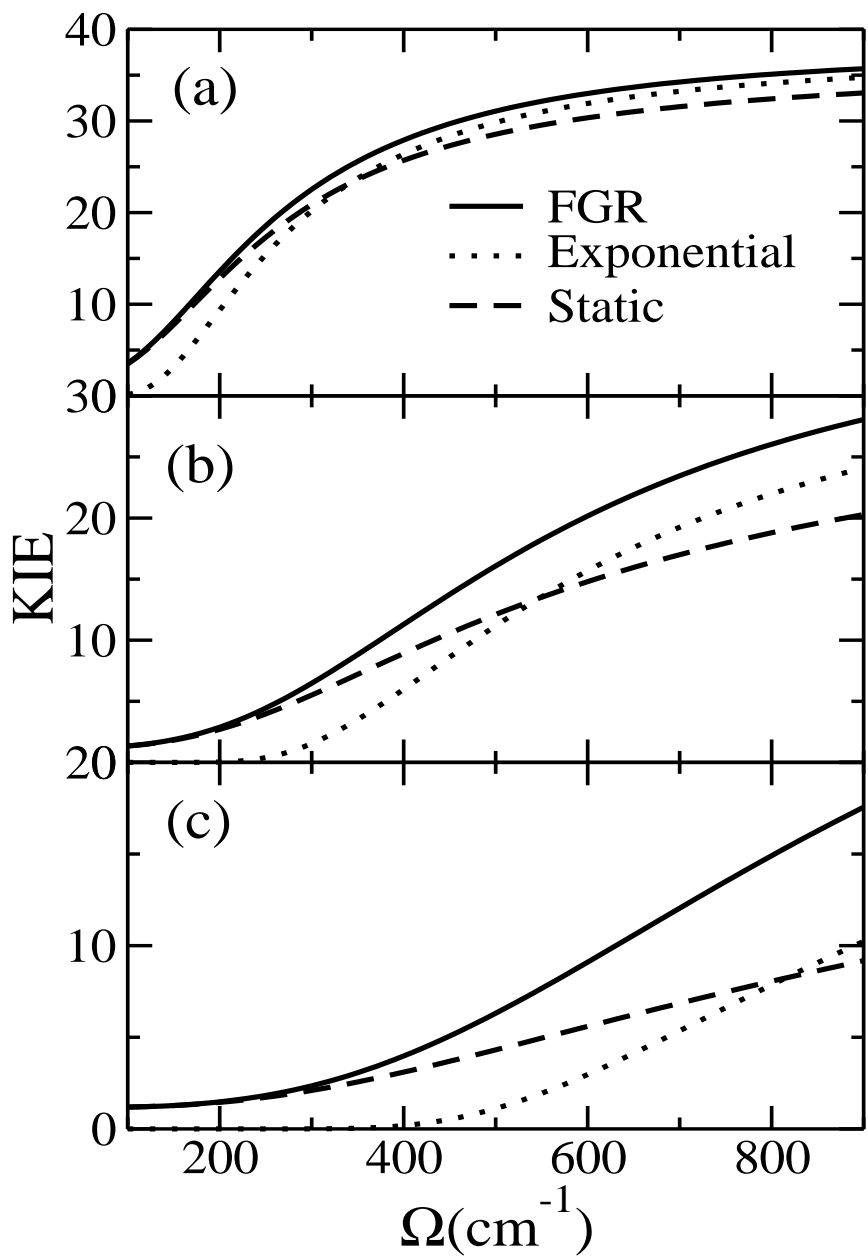

\begin{center}
\doeps{4.5in}{6.54in}{fig4}
\caption{KIE as a function of the proton donor-acceptor frequency
$\Omega$. (a) M = 100 amu, (b) M = 20 amu, (c) M = 7 amu.
The parameters are the same as in Fig. \ref{fig3}.
}
%\vspace{5em}
\label{fig4}
\end{center}
\end{figure}

Fig. \ref{fig4} plots the KIE as a function of the donor-acceptor
frequency calculated with different rate expressions. It can be seen 
that the KIEs are small at low frequencies, which means that the 
enhancement of the PCET rate due to $R$-mode fluctuations is more
significant for deuterium.
The extended UK method agrees well with the FGR result except for the 
case of high proton donor-acceptor frequencies ($\Omega > 500$ cm$^{-1}$ 
for $M$= 20 amu, and $\Omega > 400$ cm$^{-1}$ for $M$= 7 amu.).
The rate constant expression 
Eq. (\ref{eq:kmunu}) has problems in cases of small mass and frequency.
This problem is more severe when the donor-acceptor mass
$M$ is small (e.g., the left parts of Fig. 4(b) and (c))
as the exponential approximation in Eq. (\ref{eq:expr})
becomes inaccurate when the fluctuation of $R$ is large. We 
note that the failure of the exponential overlap approximation 
in calculating KIEs at low donor-acceptor frequencies has been 
discussed recently,\cite{edwards09} while it is more quantitatively 
characterized in Fig.\ref{fig4} in this study by comparing to
the exact FGR result.

In summarizing this subsection, we have shown that the model 
Hamiltonian including the DRE captures the main features 
of PCET reactions presented in many previous theoretical 
studies.\cite{cukier98,hammes-schiffer01,hammes-schiffer08,klinman06,cukier04}
The advantage of the current model is that, it 
can include the effect of the proton donor-acceptor motion
without the BO approximation for the proton motion,
as well as further approximations 
for the vibrational wave function overlap.
Our quantitative evaluation of the static $R$-mode 
and the exponential overlap approximations could also be
helpful in developing more accurate PCET theory in more 
general cases.

\subsection{Vibrational coherence at small reorganization 
energy}
Recently, abnormal temperature dependence of KIE were observed in 
experimental studies of several systems,\cite{cape05,hodgkiss06} 
where the KIE increases as the temperature increases, implying a larger
apparent activation free energy for proton transfer. Theoretical studies 
has pointed out that this could be a phenomena associated with
low reorganization energies.\cite{presse06,edwards09,ludlow09}
An interesting finding in the literature is that, when the 
reorganization energy is small (i.e., in the low friction regime), 
coherent tunneling
may play a role in the ET dynamics.\cite{onuchic88,topaler96}
It is thus interesting to investigate whether 
coherent tunneling is relevant to the abnormal temperature 
dependence of KIE in the case of small reorganization energies.

We plot in Fig. \ref{fig5} the dependence of the PCET rate constant
as a function of the driving force $\Delta G$ for $\lambda=$ 3 kcal/mol.
The other parameters are $M$=20 amu,
$\Omega=$ 300 cm$^{-1}$, and $d=$ 0.45 \AA.
The rate constants
show oscillations as a function of $\Delta G$, and this oscillatory 
behavior is more pronounced for proton than deuterium. 
This effect is readily explained in the picture of the BO 
approximation,\cite{edwards09,ludlow09,edwards09a}
with the most recent explanation presented in Ref.\cite{edwards09a}.
For example, using the extended UK type of expressions:
as the vibrational energy levels for proton motion
is 3000 cm$^{-1}$, or approximately 8.6 kcal/mol, 
the small reorganization energy
$\lambda=3$ kcal/mol is not large enough to bridge the gaps between
each pairs of vibronic transitions, and peaks for rate constants occur
at approximately $\Delta G_{\mu\nu} + \lambda \approx 0$. 
As the deuterium case has a smaller vibrational spacing, this
also explains the less oscillatory behavior for deuterium. 
Based on the same reasoning, Hammes-Schiffer and coworkers 
have also explained the abnormal temperature dependence of the KIE 
in the case of small $\lambda$.\cite{edwards09, ludlow09}

Here we provide a complementary view to the above explanation that was 
based on the BO approximation to the proton motion. 
In the FGR rate expression Eq. (\ref{eq:k10}),
the real part of the integrand 
$C_{\Delta G} (t) \equiv {\rm Re} [e^{-i\frac{\Delta G}{\hbar}t} C(t) ]$
can be regarded as the (normalized) 
reaction flux correlation function whose integral
gives the rate constant. Within first order approximation,
the time dependent acceptor 
population at short times can also be estimated as
\begin{equation}
\frac{d}{dt}P_a(t) \approx  k(t) P_d(0) \; ,
\end{equation}
where $P_a$ is the acceptor population, $P_d$ is the donor
population, and
\begin{equation}
\label{eq:szrate2}
k(t) = 2\Delta^2 {\rm Re} \int_0^t dtC_{\Delta G} (t) 
 \;.
\end{equation}

The normalized flux correlation function $C_{\Delta G} (t)$
is plotted in Fig. \ref{fig6} (a), for two different $\Delta G$ values
at -3.5 kcal/mol and -6.5 kcal/mol. The corresponding 
population evolution is shown in Fig. \ref{fig6} (b) by assuming
$P_d(0) = 1$.
We can see that due to the smaller 
damping effect associated with small $\lambda$, the second and third peaks in
the flux correlation function starts to contribute to the rate constant,
and their contribution can be either 
constructive (for $\Delta G$ = -3.5 kcal/mol, the contribution 
from second peak is 40\% from the first peak)
or destructive, (for $\Delta G$ = -6.5 kcal/mol, the contribution 
from second peak is -70\% from the first peak), 
which leads to peaks or valleys in the rate constant as shown 
in Fig. \ref{fig5}.
We note that this just the vibrationally 
coherent tunneling effect observed in the previous
numerical studies of photo-induced electron transfer reaction 
involving a high frequency intramolecular mode,\cite{topaler96} 
although a slightly different model is used in the studies
of Ref.\cite{topaler96}.

An interesting observation here is that, at $\Delta G$ values
near the peaks and valleys
of the rate constant curve in Fig. \ref{fig5}, the temperature dependence 
of the KIE shows different behavior. Fig. \ref{fig7} plots the 
KIE as a function of temperature for 
$\Delta G$ = -3.5 kcal/mol (close to a peak),
which decreases when the temperature increases;
while for $\Delta G$ = -6.5 kcal/mol (close to a valley), 
the KIE shows abnormal behavior as it increases with increasing 
temperature. The dashed lines shows the results by including only
the first peak contribution from $C_{\Delta G}(t)$, which shows 
the normal temperature dependence in both case.
Comparing the solid and dashed curves thus shows
that the contribution from the oscillations 
in $C_{\Delta G}(t)$
play an important role in the abnormal temperature dependence 
for $\Delta G$ = -6.5 kcal/mol.
This observation also holds for $\Delta G$ values near 
other peaks and valleys.
So we conclude that the abnormal temperature dependence of KIE 
in the case of small reorganization energy is  
related to coherent vibrational tunneling, which is 
a new explanation complimentary to the previous ones 
within the BO approximation of the proton motion.

\section{Conclusions}
\label{sec:conclusions}
Using a model Hamiltonian based on the Duschinsky rotation effect, 
we were able to quantitatively 
investigate the applicability of the commonly used 
BO approximation for the proton motion in studying PCET reactions. 
The extended UK method using the static $R$-mode approximation 
was found to work reasonably well except for small donor-acceptor
mass and high donor-acceptor frequencies. The exponential
approximation for the vibrational function overlap was also tested
in the model systems, and found to be inaccurate in cases of large 
proton-donor separation 
fluctuations when the proton donor-acceptor 
mass and frequency are small. This may suggest that more accurate 
analytical approximations are needed in such cases.
 
\begin{figure}
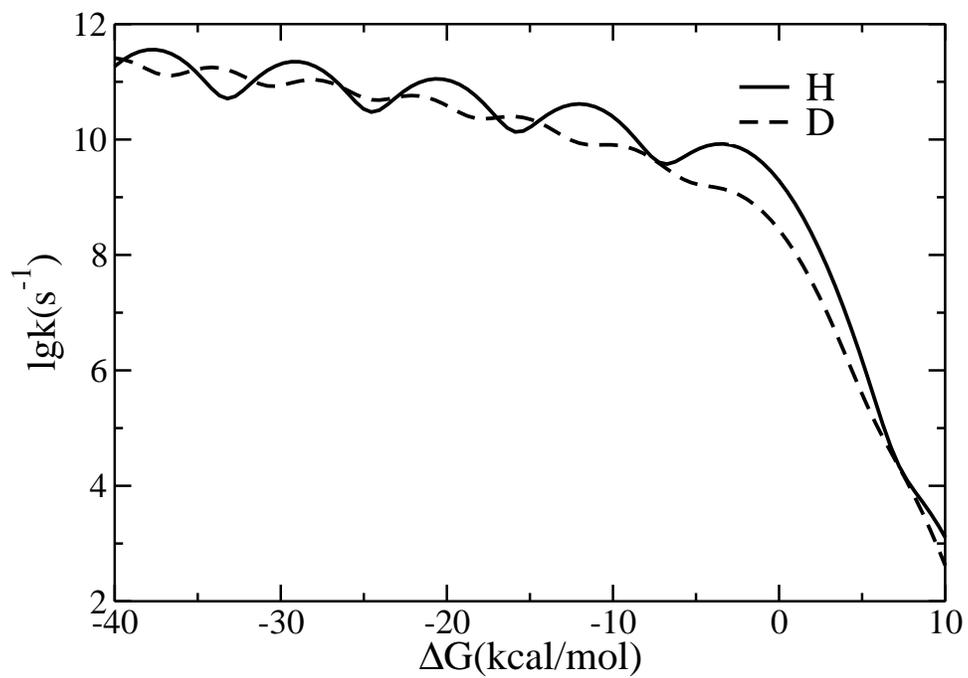

\begin{center}
\doeps{5in}{3.5in}{fig5}
\caption{PCET rate constants 
as a function of $\Delta G$ for small
reorganization energy $\lambda$ = 3 kcal/mol.
The other parameters are: 
$\Delta$ = 100 cm$^{-1}$,
$M$ = 20 amu, $\Omega$ = 300 cm$^{-1}$, 
$d$ = 0.45 \AA, and $T$ = 300 K. 
}
\vspace{5em}
\label{fig5}
\end{center}
\end{figure}

\begin{figure}
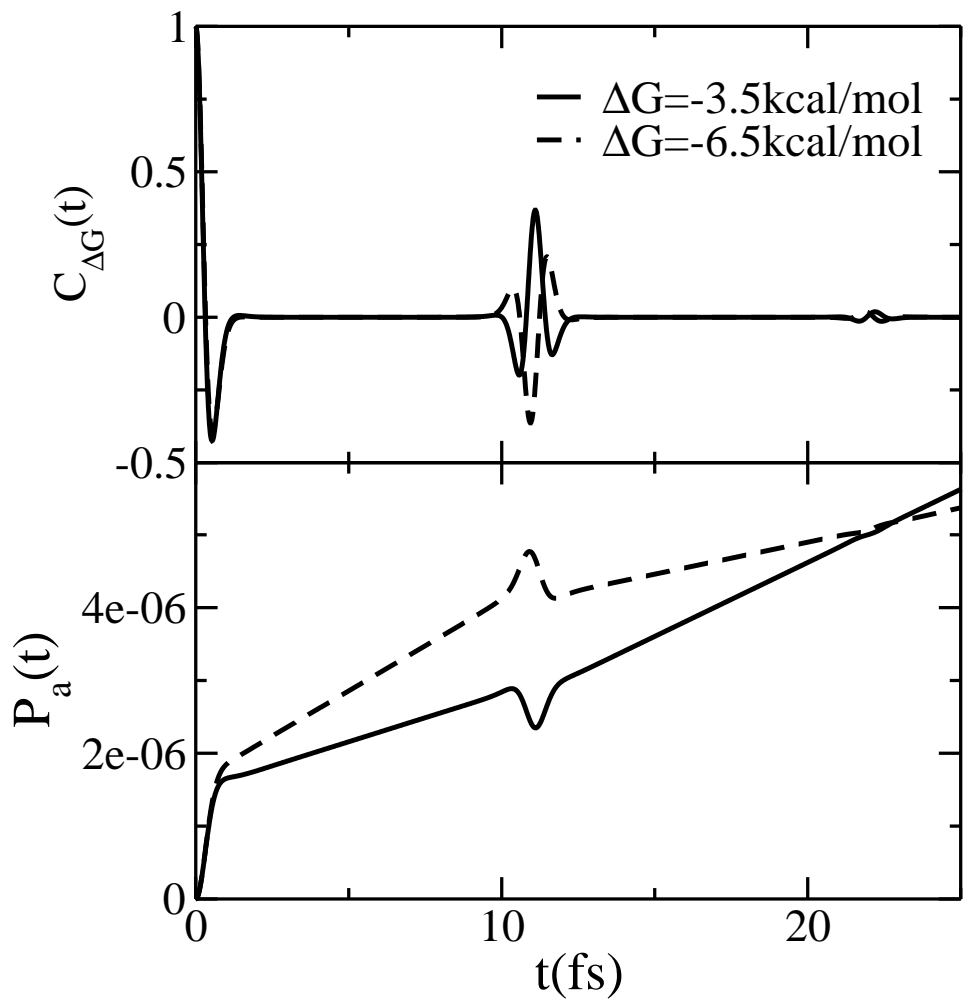

\begin{center}
\doeps{5in}{5.2in}{fig6}
%\vspace{3em}
\caption{(a) Normalized flux correlation function for 
$\Delta G$ = -3.5 kcal/mol (solid line) and -6.5 kcal/mol (dashed line). 
(b) Population of the acceptor as a function of time.
The other parameters are the same as in Fig. \ref{fig5}.
}
%\vspace{5em}
\label{fig6}
\end{center}
\end{figure}

\begin{figure}
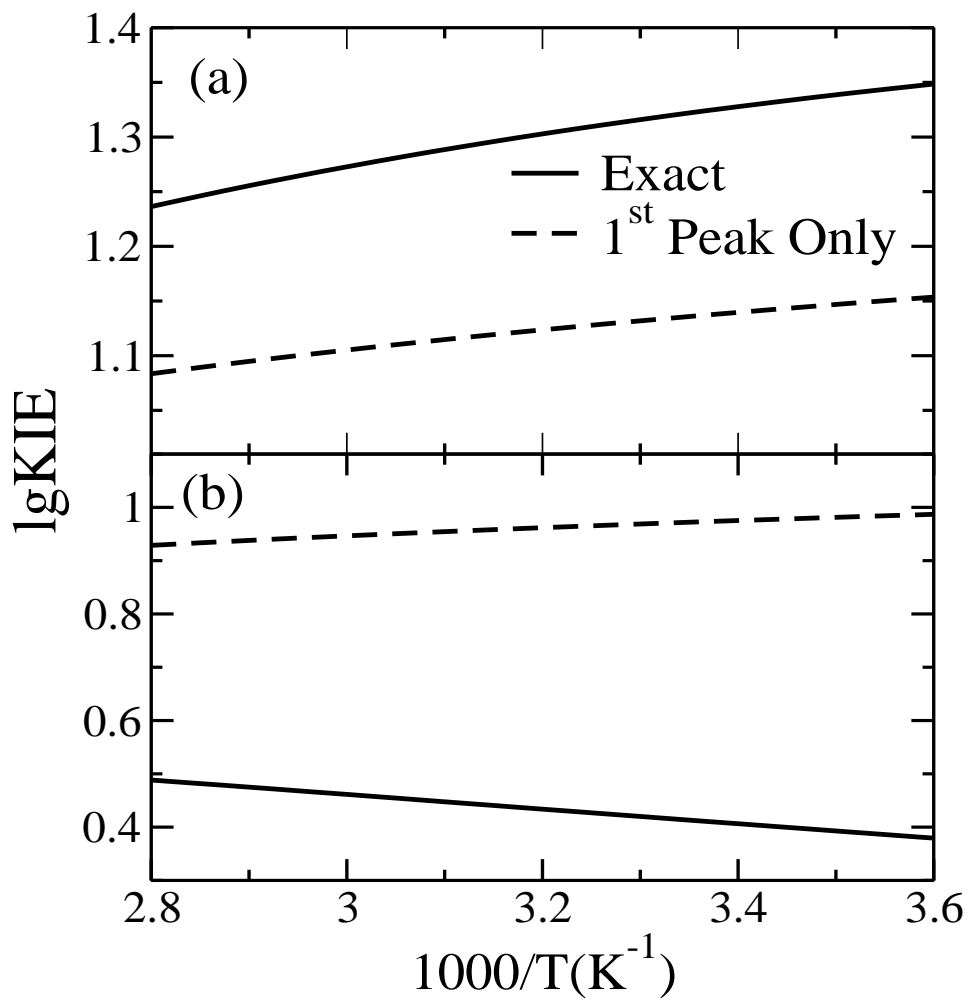

\begin{center}
\doeps{5in}{5.2in}{fig7}
\caption{KIE as a function of temperature for 
(a) $\Delta G$ = -3.5 kcal/mol, and (b) -6.5 kcal/mol. Solid lines
are the exact FGR result, while dashed lines show the results by 
including only the first peak of the correlation function 
as shown in Fig. \ref{fig5} (a).
The other parameters are the same as in Fig. \ref{fig5}.
}
%\vspace{5em}
\label{fig7}
\end{center}
\end{figure}

In the case of small reorganization energies,
we find that coherent vibrational tunneling can lead to
oscillations of rate constants with respect to the driving force 
$\Delta G$, and different temperature dependence
of KIE at $\Delta G$ near the peaks and valleys on 
the rate-driving force curve in Fig. \ref{fig5}. 
Our explanation of these results based on coherent tunnelling 
is complementary to existing theories within the 
BO approximation for
proton motion.\cite{presse06,edwards09,edwards09a,ludlow09}
As the vibrational coherent tunnelling is sensitive to 
couplings of the proton coordinate to the environmental 
DOFs,\cite{topaler96} extending the current 
study to incorporate such couplings would certainly be interesting.
The current work could also be extended to investigate photo-induced PCET 
reactions,\cite{li06,venkataraman09} where non-equilibrium dynamics
involving vibrational motion is important.

%%%%%%%%%%%%%%%%%%%%%%%%
\acknowledgments
This work is supported by
NNSF of China (20733006, 20873157, 20903101),
and the Chinese Academy of Sciences
(KJCX2.YW.H17 and the Hundred Talents Project).
%%%%%%%%%%%%%%%%%%%%%%%%

\pagebreak

\pagebreak

\pagebreak

\pagebreak

\pagebreak

\pagebreak

\pagebreak
\end{document}